\documentclass[structabstract]{aa}  
\usepackage{natbib}
\usepackage{graphicx}
\usepackage{txfonts}

\usepackage[normal]{subfigure}

\newcommand{\lsim} {\mathrel{\hbox{\rlap{\lower.55ex \hbox{$\sim$}}
                             \kern-.3em \raise.4ex \hbox{$<$}}}} 
\newcommand{\msun}{M_\odot}

\begin{document}
  \title{Systematics in lensing reconstruction: Dark matter rings in the sky?}
  \author{P.P. Ponente\inst{1,2}
    \and J.M. Diego \inst{2}
  } 
  \institute{IFCA, Instituto de F\'\i sica de Cantabria (UC-CSIC),  
    Av. de Los Castros s/n, 39005 Santander, Spain\\
    \email{ponente@ifca.unican.es}
    \and
    Departamento de F\'\i sica Moderna, Universidad de Cantabria. Av. de Los Castros s/n, 39005 Santander, Spain  \\
  }

   \date{}

\abstract 
{Non-parametric lensing methods are a useful way of reconstructing the 
lensing mass of a cluster without making assumptions about the way the mass is distributed in the cluster. These methods are particularly powerful in the case of galaxy clusters with a large number of constraints. The advantage of not assuming implicitly that the luminous matter follows the dark matter is particularly interesting in those cases where the cluster is in a non-relaxed dynamical state. On the other hand, non-parametric methods have several 
limitations that should be taken into account carefully.}  
{We explore 
some of these limitations and focus on their implications for the possible ring of dark matter around the galaxy cluster CL0024+17.}  
{We project three background galaxies through a mock cluster of known radial profile density and obtain a map for the arcs ($\theta$ map). We also calculate the shear field associated with the mock cluster across the whole field of view (3.3 arcmin). Combining the positions of the arcs and the two-direction shear, we perform an inversion of the lens equation using two separate methods, the biconjugate gradient, and the quadratic programming (QADP) to reconstruct the convergence map of the mock cluster.}  
{We explore the space of the solutions of the convergence map and compare the radial density profiles to the density profile of the mock cluster. When the inversion matrix algorithms are forced to find the exact solution, we encounter systematic effects resembling ring structures, that clearly depart from the original convergence map.}  
{Overfitting lensing data with a non-parametric method can produce ring-like structures similar to the alleged one in CL0024.}  

\keywords{gravitational lensing: strong --
  gravitational lensing: weak }

\titlerunning{Lensing systematic effects} 
\maketitle 
\section{introduction}

Gravitational lensing is one of the most powerful probes of dark matter. In particular, galaxy clusters host the strongest gravitational potentials in the Universe, hence they are rich in gravitational lensing effects. The distortions produced in the images of background galaxies by a galaxy cluster can be used to reconstruct the mass distribution of the cluster, which is believed to be largely dominated by dark matter. 
Two regimes are distinguished according to the strength of the lensing distortion. The weak lensing regime refers to small distortions that usually need to be studied in a statistical way. Large distortions, on the other hand, can be studied individually (or in pairs) and they are referred to as strong lensing. Strong lensing occurs when the projected surface mass density is on the order of the critical mass density $\Sigma_{crit}$. In this scenario, a gravitational lens bends the light in such a way that it can produce multiple images (arcs) of the same background galaxy. Each multiple image can be used as a constraint of the mass distribution. The mass distribution has to be such that, when projected back into the source plane, the multiple images concentrate (or focus) into the same point. In most cases, the number of multiple images is small, which results in few constraints. If only strong lensing is available and the number of constraints is small, one needs to rely on parametric methods. However, more and more often new data reveals large numbers of multiple images around a single cluster. The cluster A1689 is probably the most spectacular example to date where hundreds of arcs can be seen around the cluster \citep{Broadhurst_et_al_2005B,Broadhurst_et_al_2005A}. When the number of constraints is sufficiently large, non-parametric methods become competitive with the parametric ones and with the advantage that no {\it a priori} assumption is made about the mass distribution of the cluster. Non-parametric methods applied to lensing mass reconstruction have been studied in the past \citep{Saha_Williams_1997,Abdelsalam_et_al_1998, Bridle_et_al_1998, seitz_et_al_1998,Kneib_et_al_2003,Diego_et_al_2005a,Diego_et_al_2005b,Smith_et_al_2005, Bradac_et_al_2005, Halkola_et_al_2006, Cacciato_et_al_2006}). On the positive side, in cases where the number of constraints is large, the results obtained with the parametric and non-parametric methods agree well (Diego et al. 2005b) probing, among other things, that the dark matter does trace the luminous matter and the usefulness of non-parametric methods as a way of testing that the assumptions made in the parametric methods are well founded. Non-parametric methods have been used as well to combine weak and strong lensing data in the same analysis \citep{Abdelsalam_et_al_1998b,Bridle_et_al_1998,Saha_Williams_Abdelsalam_1999,Kneib_et_al_2003,Smith_et_al_2005,Bradac_et_al_2005,Diego_et_al_2007}.
 
On the other hand, non-parametric methods have a series of limitations. In this paper we explore one of these limitations related to the limited resolution in the mass reconstruction and its connection with the accuracy in the reconstructed arc positions.\\

The results of this paper may have implications for the results of \citet{Jee_et_al_2007}, who use a non-parametric method and find an unusual ring of dark matter around the cluster. While we do not question the validity of these interesting results, we explore the possibility that spurious structures might appear when using non-parametric methods if the limitations of parametric methods are not taken into account in the analysis.

\subsection{A ring of dark matter around CL0024+17?}
The cluster CL0024+17 ($z=0.395$) was one of the first for which strong lensing was observed \citep{Wallington_et_al_1992}. Four strongly lensed arcs can be clearly seen around the tangential critical curve \citep[][see also]{Smail_et_al_1996,Broadhurst_et_al_2000}. These arcs have been used to constrain the mass in the central region of the cluster \citep{Colley_et_al_1996, Tyson_et_al_1998, Broadhurst_et_al_2000, Comerford_et_al_2006}. These mass constraints have been compared with those derived from X-ray measurements with CHANDRA \citep{Ota_et_al_2004} and XMM-Newton \citep{Zhang_et_al_2005}. These authors estimated that the X-ray masses are a factor 3-4 lower than the lensing masses. This discrepancy has been interpreted as a sign that the cluster is not in  hydrostatic equilibrium.

In \citet{Jee_et_al_2007}, the authors reconstruct the mass of the cluster out to 100 arcseconds from its center. This corresponds to a physical size of $0.389$ Mpc for an object located at $z \simeq 0.4$. In their analysis, they combine strong and weak lensing with a non-parametric method. The authors find a dark matter ring surrounding the cluster core, at $r \approx 75$ arcseconds from the center \citep[Fig.~10 in][]{Jee_et_al_2007}. The authors suggest that this ring might be the result of a high speed collision between two clusters along the line of sight \citep{Czoske_et_al_2001} in an scenario similar to the 'bullet cluster' \citep{bullet_cluster} but with the difference that in that case the collision is perpendicular to the line of sight.\\

Whether the existence of the dark matter ring is real or not has been debated by many other authors \citep{Milgrom_Sanders_2008,Qin_et_al_2008,ZuHone_et_al_2009,Zitrin_et_al_2009,Umetsu_et_al_2010}. \citet{Milgrom_Sanders_2008} reconstruct the radial profile of the mass assuming a model based on modified Newtonian dynamics (or MOND). The authors claim that a ringlike structure appears at the MOND transition region (see figs. 3 and 4 in their paper). According to the authors,  CL0024 can be considered as a robust probe of MOND. In \citet{Qin_et_al_2008}, the authors study the distribution of galaxies in CL0024, which, being collisionless, should exhibit a similar ring-like pattern. On the basis of 295 counts, the authors find no evidence of a ring in the distribution of galaxies. In a different paper, \citet{ZuHone_et_al_2009} use a hydrodynamical simulation of two collisioning clusters to compute the radial profiles after the collision. They find no evidence of either a dip or ring in the radial profile outside the core radius after the collision. They conclude that a ring-like feature could only be explained by an unlikely and highly tuned set of initial conditions before the collision.

To reanalyze the lensing data for CL0024, \citet{Zitrin_et_al_2009} analyze this cluster using data from the Hubble Space Telescope (HST) instrument ACS/NIC3. The dark matter distribution profile was reconstructed using a SL parametric method based on six free parameters. The results presented in Fig. 1 and Fig. 2 of their paper reveal neither a  dip nor ring in the profiles. Finally, \citet{Umetsu_et_al_2010} combine a large field of view data set from the SUBARU telescope with data from HST ACS/NIC3, finding no evidence of the ringlike structure after the mass reconstruction (see Fig. 21 of their paper).\\

In this paper, we revisit the debate using a non-parametric method similar to that used in \citet{Jee_et_al_2007} but applied to simulated data (weak and strong lensing). The advantage of using simulations is that the underlying dark matter distribution and the position and redshifts of the background sources are perfectly known. This offers the unique possibility of comparing the optimal solution with the multiple possible solutions obtained by the non-parametric method. We can also explore the space of solutions obtained when the minimization is done under different assumptions and compare with the original mass distribution. \\ 

In Sections \ref{sectII} and \ref{sectIII}, we introduce the fundamentals of the gravitational lensing and the non-parametric method used in this paper for the mass reconstruction. In  Section \ref{sectIV}, we describe the mock data used in our analysis. In Section \ref{sectV}, we present the results obtained by our non-parametric method and compare the different solutions with the optimal one. Finally, in Section \ref{sectVI}, we discuss our results and in Section \ref{sectVII} our conclusions.\\

\section{Gravitational lensing basics} \label{sectII}  
In gravitational lensing, it is usual to adopt the thin lens approximation because the cosmological distances between the observer, the lens, and the sources are much greater than the size of the lens.~Hence, the lens can be treated as a plane. All the other elements in the lensing problem are also assumed to be located in planes. When there are multiple background galaxies, each one is assumed to be in a different plane with redshift $z_i$ (in the case of strong lensing) or in the same plane at the average redshift $z$ (in the case of weak lensing). All these planes are perpendicular to the line of sight and the deflection is assumed to occur instantly when the light crosses the lens plane. \\
We define $D_{ls}$ as the angular diameter distance between the source plane and the lens plane and $D_{ol}$ and $D_{os}$ as the angular diameter distances from the observer to the lens and from the observer to the sources, respectively. With respect to the line of sight, the sources are located at angular positions ${\bf \beta}_i$ ($i=1,2,...,n$ with $n$ the number of sources), while the lensed images are located at positions ${\bf \theta}_i$ ($i=1,2,...,m$ with $m$ the number of images). We define the equation of the lens

\begin{equation}\label{lens_equation}
\beta = \theta - \frac{D_{ls}}{D_{os}}\alpha(\theta).
\end{equation}
We denote by $\psi(\theta)$ the two-dimensional potential produced by all the masses located at $\theta '$
\begin{equation}\label{2-dim_potential}
\psi(\theta) = \frac{4 G D_{ol}D_{ls}}{c^2 D_{os}} \int d^2\theta' \Sigma(\theta')ln(|\theta - \theta'|),
\end{equation}  
where $\Sigma(\theta')$ is the surface density of the cluster at the given position $\theta'$. The part outside the integral is related to the {\it critical density}
\begin{equation}\label{sigma_crit}
\Sigma_{crit} \equiv \frac{c^2}{4\pi G} \frac{D_{os}}{D_{ol} D_{ls}}.
\end{equation}
The above equation is used in the definition of the {\it convergence}
\begin{equation}\label{convergence}
\kappa = \frac{\Sigma(\theta)}{\Sigma_{crit}}
\end{equation} 
The deflection angle $\alpha$ and the convergence can be expressed as derivatives of the two-dimension potential
\begin{equation}\label{alfa_psi}
\alpha  =  \nabla \psi, 
\end{equation}
\begin{equation}\label{kappa_psi}
\kappa  =  \frac{1}{2}\nabla^2 \psi. 
\end{equation}
The magnification that the lens produces on the source is quantified by the determinant of the matrix describing the variation in the image position $\delta \theta$ for a small variation in the source position $\delta \beta$
\begin{equation}\label{small_variation}
\mu = \det \left | \frac{\partial \theta}{\partial \beta} \right | = \left [ \det \left | \frac {\partial \beta}{\partial \theta} \right | \right ]^{-1}. 
\end{equation}
From Eq. (\ref{lens_equation}), we get
\begin{equation}\label{magnification}
\mu^{-1} = 1 - \frac{\partial \alpha_x}{\partial \theta_x} -\frac{\partial \alpha_y}{\partial \theta_y} + \frac{\partial \alpha_x}{\partial \theta_x} \frac{\partial \alpha_y}{\partial \theta_y} - \frac{\partial \alpha_x}{\partial \theta_y} \frac{\partial \alpha_y}{\partial \theta_x}. 
\end{equation} 

The strong lens regime is most sensitive to the central mass of the cluster, where the mass surface density is normally higher than the critical surface mass density ($\kappa > 1$). When the surface mass density drops significantly below the critical density ($\kappa<<1$), we are in the regime of weak lensing. Weak lensing cannot produce multiple images, but useful information about the distribution of the mass in the cluster can be extracted from the {\it shear} of the distortion ($\gamma_1$ and $\gamma_2$). Differentiating Eq. (\ref{lens_equation}), we obtain
\begin{equation}\label{partial_potential}
H = \delta_{ij} - \frac{\partial \psi}{\partial \theta_i \partial \theta_j} = \left ( 
\begin{array}{cc}
1- \kappa - \gamma_1 & - \gamma_2 \\
- \gamma_2 & 1 - \kappa + \gamma_1 \\
\end{array}
\right )
\end{equation}
where
\begin{equation}
\gamma_1(\theta)  =  \frac{1}{2}(\psi_{11}-\psi_{22}),   
\label{gamma1}
\end{equation}
\begin{equation}
\gamma_2(\theta)  =  \psi_{12}=\psi_{21},
\label{gamma2}
\end{equation}
where the double subscripts indicate the second order partial derivative. Equations (\ref{gamma1}) and (\ref{gamma2}) can be expressed in the complex notation 
\begin{equation}\label{complex_shear}
{\bf \gamma} = \gamma_1 + {\bf i} \gamma_2
\end{equation}
to obtain the amplitude and the orientation of the deformation. The {\it reduced shear} is defined (in complex notation) ${\bf g} = {\bf\gamma}/(1-\kappa)$. The shear measures coherent shape distortions of source galaxies.\\
The detection of multiple images and/or the measurement of the shear can be used to constrain the mass distribution of the cluster. In cases where the number of constraints is large, the mass of the cluster expressed in Eq. (\ref{2-dim_potential}) can be reconstructed using a non-parametric method.

\subsection{Parameter-free lensing reconstruction }
Here we adopt formalism and notation of \citet{Diego_et_al_2005a} and \citet{Diego_et_al_2007}.
 
The mass reconstruction described in those papers is based on a parameter-free method where the lens plane is divided into a finite number of cells $N_c$ and Eq. (\ref{lens_equation}) can be written in algebraic form. The deflection angle $\alpha$ at a position $\theta$ is computed from the net contribution of the discretized mass distribution $m_i$ at the positions $\theta_i$
\begin{equation}\label{discrete_alfa}
\alpha(\theta) = \frac{4 G}{c^2} \frac{D_{ls}}{D_{os} D_{ol}} \sum_{N_c} m_i(\theta_i)\frac{\theta - \theta_i}{| \theta - \theta_i |^2}.
\end{equation}
The number of cells in the gridded mass must be carefully choosen. The discretization of the lens plane affects the spatial resolution of the mass reconstruction, as we discuss in more detail later.

All the positions of the pixels hosting a strong lens image can be described by the vector $\theta$ of dimension $N_{\theta}$. For each pixel in the $\theta$ vector and for a given discretized mass distribution, a corresponding $\beta$ pixel can be traced back to the source plane. The relation between all these elements can be written in algebraic form 
\begin{equation}\label{lens_matrix_1}
\theta = \Upsilon M + \beta,
\end{equation}    
where $\theta$ (and $\beta$) are vectors containing the $x$ and $y$ components of the $N_{\theta}$ pixels of the arcs (and sources), $M$ is the vector of the masses inside the $N_c$ cells, and the matrix $\Upsilon$ has the dimension of $(2N_\theta \times N_c)$. The description of this matrix is given in \citet{Diego_et_al_2005a}.

Eq. (\ref{lens_matrix_1}) is a system of $2N_{\theta}$ linear equations whose solution can be achieved using the methods described in Diego et al. (2005a). The unknowns of the problem are the masses in the $M$ vector and the central positions of the background sources. Both vectors can be united into a single one $X$, rendering the simpler equation
\begin{equation}\label{lens_matrix_2}
\theta = \Lambda X ,
\end{equation}
where $\Lambda$  is a matrix similar to $\Upsilon$ but with an extra sparse block containing $1$ and $0$.\\  

Weak lensing data can be modeled in a similar way. 
The two components of the shear are computed through the matrices that represent the contribution of each mass cell:

\begin{equation}\label{weak_matrix}
\left(
\begin{array}{c}
\gamma_1 \\
\gamma_2 \\
\end{array}
\right) = \left(
\begin{array}{c}
\mathbf{\Delta_1} \\
\mathbf{\Delta_2} 
\end{array}
\right) M.
\end{equation}
A detailed description of the matrices $\Upsilon$, $\Delta_1$ and $\Delta_2$ is presented in Appendix A.

After including the weak lensing regime, the joint system of linear equations can be explicitly written down as
\begin{equation}\label{lens_matrix_3}
\left( 
\begin{array}{c}
\theta_x \\
\theta_y \\
\gamma_1 \\
\gamma_2 \\
\end{array} \right)= \left(
\begin{array}{ccc}
\mathbf{\Upsilon_x} & \mathbf{I_x} & \mathbf{0} \\
\mathbf{\Upsilon_y} & \mathbf{0} & \mathbf{I_y} \\
\mathbf{\Delta_1} & \mathbf{0} & \mathbf{0} \\
\mathbf{\Delta_2} & \mathbf{0} & \mathbf{0} \\
\\
\end{array}
\right) \left(
\begin{array}{c}
M \\
\beta_x \\
\beta_y \\
\end{array}
\right),
\end{equation}
where the element {\it ij} in the matrix $I_x$ is 1 if the $\theta_i$ pixel comes from the $\beta_j$ source, and is 0 otherwise. The matrix $0$ is the null matrix. Eq. (\ref{lens_matrix_3}) can be written in the more compact form
\begin{equation}\label{lens_matrix_4}
\Phi = \mathbf{\Gamma} X,
\label{main_system}
\end{equation}
where $\Phi$ is the vector containing the positions of the arcs and the shear measurements, $\Gamma$ is a non-square matrix, and $X$ is the vector of the unknowns. \\

Written in this simple form, the lensing problem could, in principle, be resolved after the inversion of Eq. (\ref{lens_matrix_4}), $X = \Gamma^{-1}\Phi$.\\

\section{Inversion of the lens equation} \label{sectIII}
The vector $X$ can be found by inverting Eq. (\ref{lens_matrix_4}). However, the matrix $\mathbf{\Gamma}$ is often non-invertible. This is actually not a problem as we seek an approximate solution with a more physical meaning than the exact solution. One of the assumptions made in the parametric method is that the background galaxies are infinitely small. The exact solution of the system of linear equations would reproduce an unphysical situation where the background galaxies are {\it point-like}. On the other hand, an approximate solution of the system has the benefit that the predicted background sources are not point-like but extended. In addition, an approximate solution allows for some error that is needed to compensate for the other {\it wrong} assumption made in non-parametric methods, namely, the assumption that the mass distribution is discretized. The predicted size of the background sources can be controlled in the solution by setting an error level or {\it residual}, $R$, in the system of linear equations
\begin{equation}
\mathbf{R} \equiv  \Phi - \mathbf{\Gamma} X.
\label{residual}
\end{equation}
In the case of WL, the physical meaning of the residual is the associated error in the determination of the reduced shear.
 
As discussed in Diego et al. (2005a), a powerful way to find an approximate solution to the system is through the bi-conjugate gradient algorithm, which minimizes the square of the residual 

\begin{eqnarray}\label{res_square}   
&& \mathbf{R^tC^{-1}R}  = (\mathbf{\Phi} -\mathbf{\Gamma X})^t\mathbf{C^{-1}}(\mathbf{\Phi} -\mathbf{\Gamma X})  \nonumber \\   
&&\  =  \left( \mathbf{\Phi}^t\mathbf{C^{-1}}\mathbf{\Phi} -2 \mathbf{\Phi}^t\mathbf{C^{-1}}\mathbf{\Gamma X} +\mathbf{X}^t \mathbf{\Gamma}^t\mathbf{C^{-1}}\mathbf{\Gamma X} \right),
\end{eqnarray}
where $\mathbf{C}$ is the covariance matrix of the residual $R$ and among other things includes the relative weights of the SL and WL data. As discussed in \citet{Diego_et_al_2007}, this residual can be described (to first order) by a Gaussian distribution with a diagonal covariance matrix. This is however an approximation. The elements of the residual are correlated with each other, in particular those elements corresponding to the SL part of the data. The elements of the WL part of the residual are far more weakly correlated with each other and the diagonal approximation is a far more valid for this part. For the time being, we assume that the covariance matrix is diagonal and later discuss its implications. The diagonal approximation has been also assumed in previous works, including \citet{Jee_et_al_2007}.  The elements of the diagonal corresponding to the SL data are set to $\sigma_{SL}$ and the elements of the diagonal corresponding to the WL data are set to $\sigma_{WL}$. We adopt $\sigma_{SL} \sim 1$ arcsecond (in radians) and  $\sigma_{WL} = 0.3$ (or equivalently 30\%). As discussed in \citet{Diego_et_al_2007}, the value of $\sigma_{SL}$ has a physical meaning. Its value is connected with the angular size of the sources.

An alternative to the bi-conjugate gradient is the non-negative quadratic programming (QADP). A brief description of bi-conjugate and quadratic programming is given in Appendix B.

Both methods have advantages and disadvantages: the bi-conjugate gradient is extremely fast, although the final solution may contain unphysical negative masses. On the other hand, the non-negative quadratic programming algorithm does not produce a solution with negative masses, but it is significantly slower than the bi-conjugate gradient (its typical computation time is a few hours compared with a few minutes to reach similar accuracy). In both cases, a threshold $R^2 \approx \epsilon$ is defined to set the level at which the minimization stops. 

The method has one drawback when applied to our problem: one can not choose $\epsilon$ to be arbitrary small. If one chooses $\epsilon$ to be very small, the algorithm will try to find a solution that focuses the arcs into $N_{\rm s}$ sources with unphysically small sizes. The mass distribution that accomplishes this, is usually very biased relative to the correct one: it usually has a lot of substructure with large mass fluctuations in the lens plane. One must then choose $\epsilon$ with some carefully selected criteria. Since the algorithm will stop when $R^2 < \epsilon$, we should choose $\epsilon$ to be an estimate of the expected error associated with the sources not being point-like and the reconstructed mass being discretized. Instead of defining $\epsilon$ in terms of $R^2$, the parameter $\epsilon$ should be defined in terms of the residual of the conjugate gradient algorithm $r_{\rm k}$ (see Eq. \ref{res_step} in Appendix B). This would accelerate the minimization process significantly since we would not need to calculate $R$ at each step but use the already estimated $r_{\rm k}$. Both residuals are connected by the relation
\begin{equation}
r_{\rm k} = \Gamma^{\rm T}R.
\end{equation}
Imposing a prior on the size of the sources means that we expect the residual of the lens equation, $R$, to take typical values on the order of the expected dispersion (or size) of the sources at the measured redshifts. Hence, we can define a $R_{\rm prior}$ of the form
\begin{equation}   
R_{\rm prior}^i = \sigma_{\rm prior}^{\rm i} * RND,
\end{equation}
where the index $i$ runs from 1 to $N_\theta$ and $\sigma_{\rm i}$ is the dispersion (prior) assumed for the source associated with pixel $i$ and $RND$ is a random number normally distributed with zero mean and unity variance. We can then estimate $\epsilon$ as 

\begin{equation}\label{epsilon}
\epsilon = r_{\rm k}^{\rm T}r_{\rm k}=R_{\rm prior}^T \Gamma \Gamma^{\rm T} R_{\rm prior}.
\end{equation}

Following Diego et al. (2005a), we construct $R_{\rm prior}$ assuming that the source galaxies can be described as Gaussians with $\sigma = 30h^{-1}$ kpc. In our particular problem (a grid with $N_{\rm c} = 32 \times 32$ cells), this results in a value $\epsilon \approx 2 \times 10^{-10}$.  One has to be careful not to choose a too small $\sigma$. They should be larger than the typical size of a galaxy. Only when the number of grid points, $N_{\rm c}$, is large enough, can the gridded version of the real  mass distribution focus the arcs into sources that are similar in size to real ones. If $N_{\rm c}$ is not large enough, the gridded version of the true mass focuses the arcs into sources that are larger than the real sources. This is explained in more detail below.\\
 
The choice of the threshold is a crucial point when performing the mass reconstruction. We illustrate in the next few sections how this affects both the final mass estimation and the positions of the sources.

\section{Simulation of mock lensing data} \label{sectIV}        
We now describe the simulated data consisting  of a simple cluster and lensing (both strong and weak) data set. The use of simulated data gives us the unique advantage of being able to compare the reconstructed mass with the true underlying simulated mass and check for biases and systematics.

\begin{figure}
  \centering
  \subfigure{\includegraphics[width=80mm]{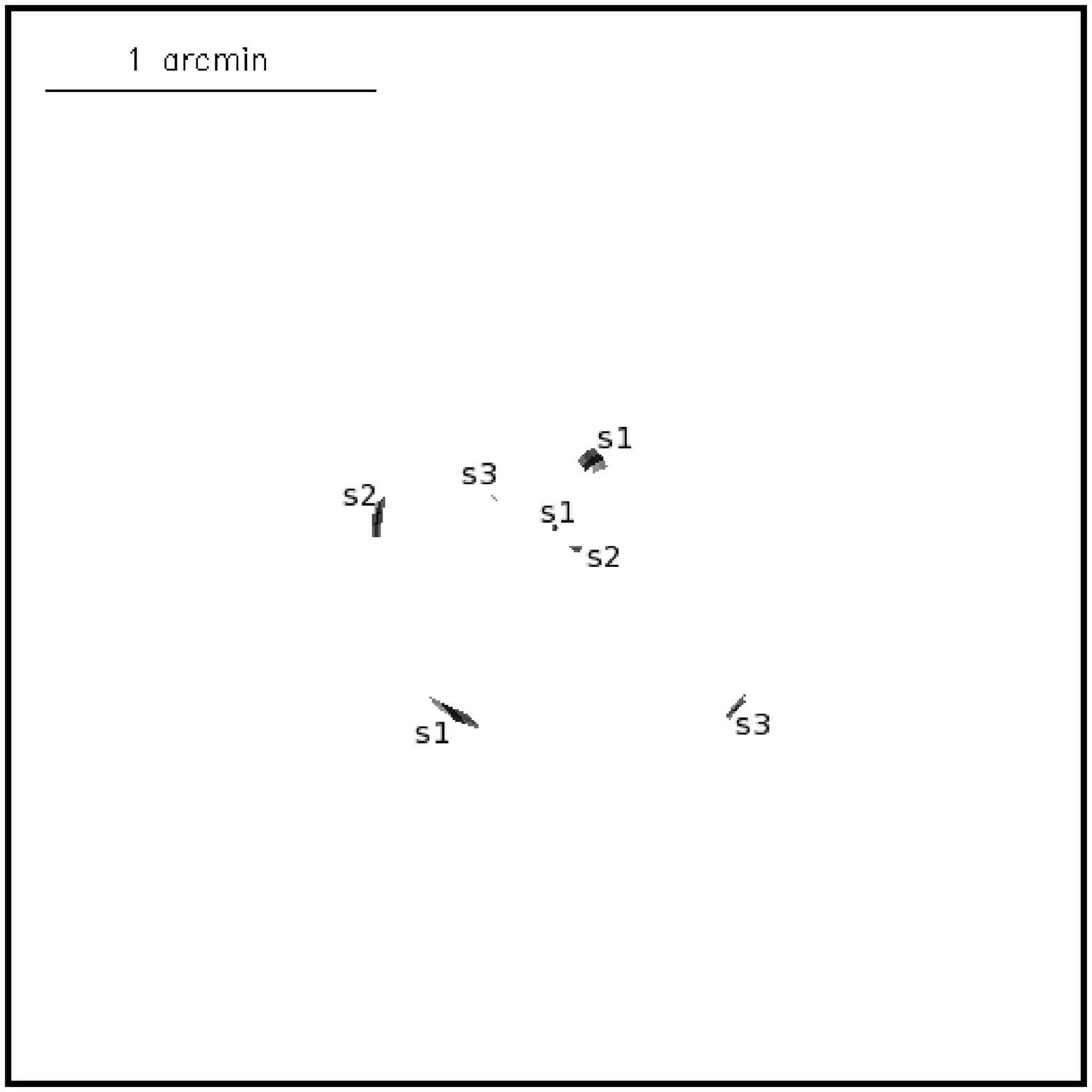}} 
  \subfigure{\includegraphics[width=80mm]{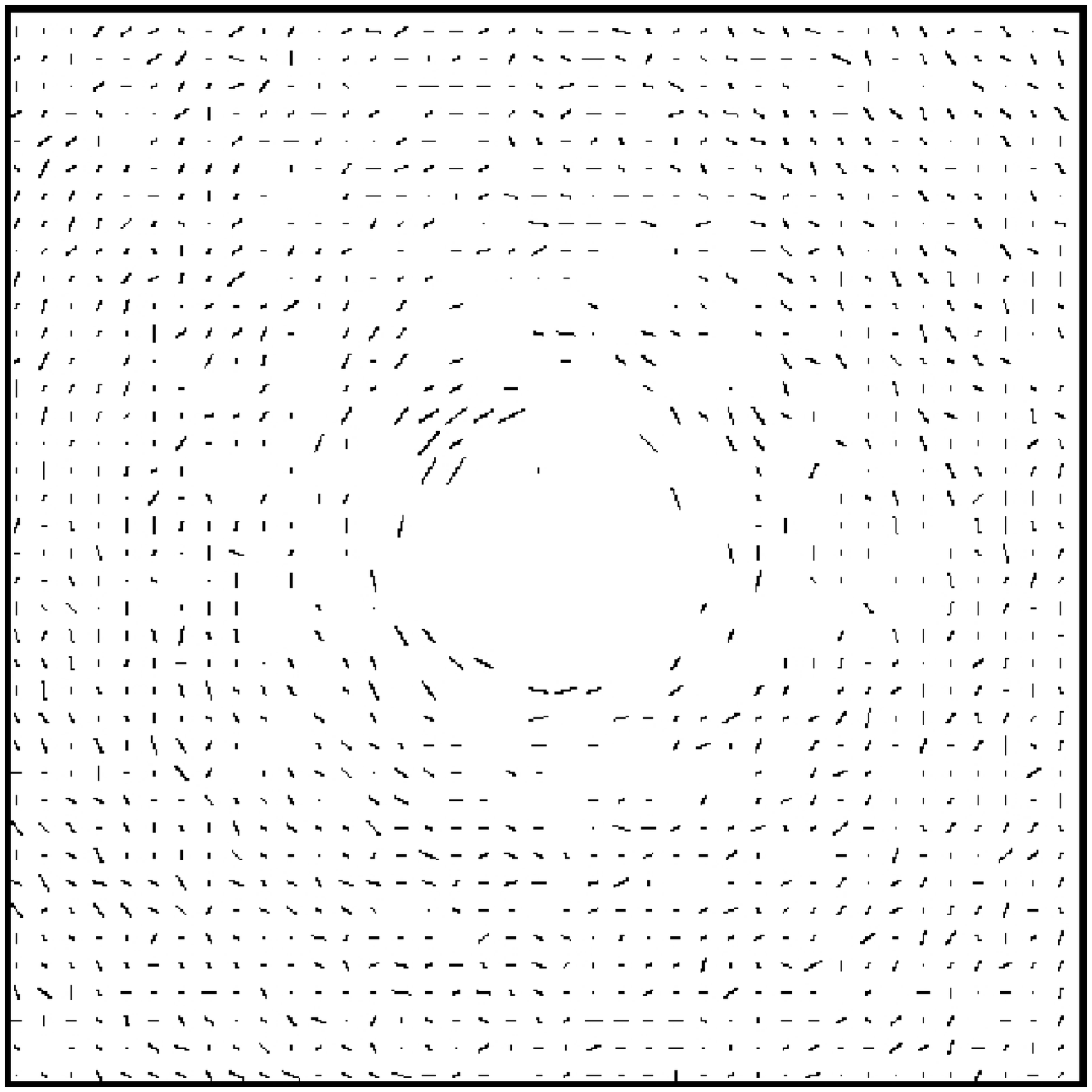}} 
  \caption{{\it Top panel:} The lensed arcs ($\theta$ map) originated from three sources in the background (not shown in the figure). 
    The total number of pixels forming the arcs is $N_\theta = 288$. 
    {\it Bottom panel:} shear field derived from the lens and used for the weak lensing computation. The inner points have been removed to mimic the contamination from cluster member galaxies. Total number of shear points is $N_{\rm shear} = 1301$, needed to set the dimension of the lensing matrix. All points have a Gaussian noise of 30\%.}
  \label{cluster_grid_shear}
\end{figure}

For the cluster, we assume a single Navarro-Frenk-White  \citep[NFW,]{NFW} profile for the radial density. We choose the simplest possible profile in order to avoid the effects of the uncertainties caused by the complexity of the mass distribution. We also assume the same redshift of CL0024 ($z=0.4$), while the field of view corresponds to the field of view of the ACS field (FOV=3.3 arcmins). The resulting mass in the whole field of view is $M(\mbox{FOV}) \sim 4.8 \times 10^{14} \msun$, while  when we consider core radii within 30'', we have $M(<30'') \sim 1.28 \times 10^{14} M_\odot$ (the mass reconstruction in \citet{Jee_et_al_2007} yields $M(r <30'')\approx (1.79 \pm 0.13) \times 10^{14}M_\odot$).\\
The strong and weak lensing data are computed using the full resolution of our simulated cluster (in the reconstruction process, the lens plane is divided with a grid that effectively reduces this resolution).\\ 
For the strong lensing data, we assume the same number of background sources ($N_s=3$) identified in \citet{Jee_et_al_2007} and that their redshifts are $z_1 = 1.675$, $z_2 = 1.27$, and $z_3 = 2.84$. We carefully chose the position of the background sources in trying to mimic the strong lensing data set used by \citet{Jee_et_al_2007}, although this is not really relevant to our work.
They identify five arcs from source 1, two arcs from source 2 and two arcs from source 3, making a total of nine. Most of the arcs are tangential, particularly those originating from source 1, which indicates that this source has to be positioned very close (in projection) to the density peak of the lens. In our case, our simulated strong lensing data set consists of seven arcs, three of which originate from source $s_1$ (two tangential and one radial), two from source $s_2$ (one tangential and one radial), and  two from source $s_3$ (one tangential and one radial). The map of the lensed images is represented in Fig.~\ref{cluster_grid_shear} (top panel), with the labels identifying the original sources.

The shear data is computed assuming that the density of available background galaxies is lower toward the center of the cluster, where the presence of the cluster itself makes it harder to estimate the reduced shear. For all the shear data points, we assume a Gaussian noise of 30\%. In addition to the cluster itself, the magnification bias has to be taken into account. Magnification acts on galaxies (enhancing their flux) but also expanding the area of the sky behind the cluster. In \cite{Broadhurst_et_al_2005A}, the latter effect is estimated and showed that a net deficit of background galaxies is expected \citep[see also][]{Umetsu_et_al_2011}. The resulting shear field is shown in Fig.~\ref{cluster_grid_shear} (bottom panel).

\subsection{Simulated vs real data}\label{Sect_4_1}
In \citet{Jee_et_al_2007}, the authors consider a FOV of $3.5 \times 3.5$ arcminutes that is gridded in a $52 \times 52$ regular grid, but with the four corner points removed. We consider a slightly smaller FOV ($3.3 \times 3.3$ arcminutes) and divide the FOV using a $32 \times 32$ regular grid. We chose the side of the grid to be 32 to ensure that the number of constraints is comparable to the number of unknowns and hence have a more stable system of equations. A  larger number of grid points will only introduce unnecessary noise in the reconstructed solution.

In \citet{Jee_et_al_2007}, the strong lensing constraints are derived from 132 {\it knots} identified in the lensed images and the weak lensing constraints are based on an ensemble of 1297 background galaxies with photometric redshifts $z_{\rm phot} \geq 0.8$. In our simulated data, we instead consider all the pixels of our lensed images (288 pixels) for the strong lensing, while for the weak lensing we create a simulated vectorial field in 1301 positions. \\
The solution in \citet{Jee_et_al_2007} is found after a minimization process involving the strong and weak lensing data, a regularization term and a model for the lensing potential. The regularization term improves the smoothness of the recovered solution and in principle helps to reduce the overfitting problem. The method is based on the maximum entropy method (MEM), which has a positive prior that forces the improved solution to remain positive. Here, we also use a minimization process but instead of a regularization term we stop the minimization process at a point that avoids overfitting the data. An interesting discussion of this point can be found in \citet{Jee_et_al_2007}. They perform a {\it delensing} of the arcs from one particular source. The resulting recovered sources are reported in Fig. 14 in their paper, where the orientation, parity, and size of the images are strongly consistent among the different recovered sources. Nonetheless, the positions of the the delensed images do not overlap. The same authors report: {\it ' When we forced the two locations to coincide in our mass reconstruction, the smoothness of the resulting mass map was compromised'}. This might indicate a tension between the recovered solution and the corresponding goodness of fit. Formally the solution is not an optimal one in the sense that the recovered source positions do not coincide but seem to be good enough to ensure that the recovered sources resemble the real ones.

\begin{figure}
\centering
\includegraphics[width=80mm]{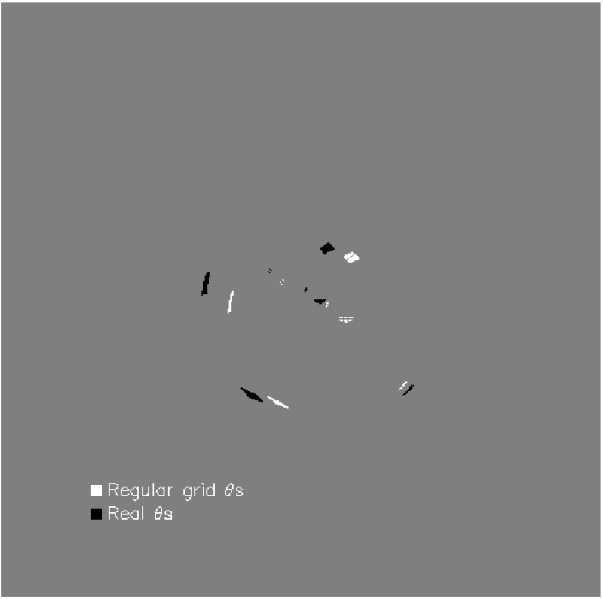}
  \caption{Simulated observed arcs (black) versus predicted ones from the optimal solution (white). The difference between the two sets of $\theta$ positions is representative of the error expected when recovering the solution.}
  \label{arcs2}
\end{figure}

\section{The optimal solution} \label{sectV}
With the simulated data, a very interesting exercise can be done before attempting the mass reconstruction. Since we know the true underlying mass and the positions of the background sources, we can predict where the arcs should appear when we assume the {\it optimal} solution possible for $X$ assuming a uniform grid with $32\times32$ cells. This solution consists of the mean mass in each cell corresponding to the true underlying mass and the three real positions.  
\begin{figure}
  \centering
  \subfigure{\includegraphics[width=80mm]{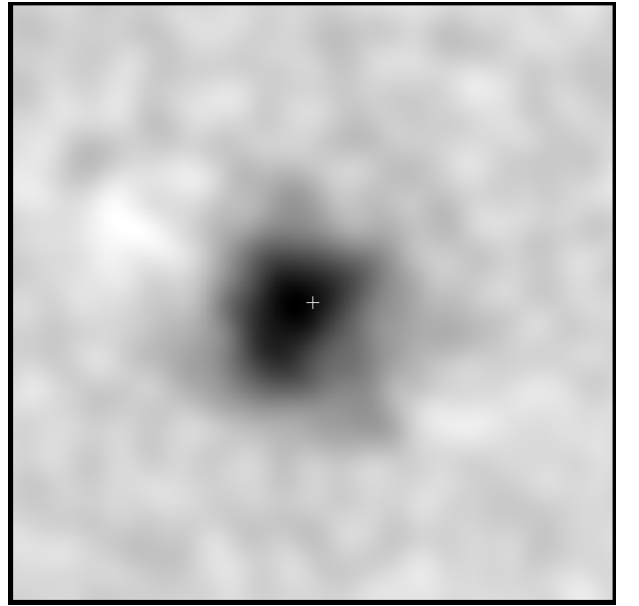}}
  \subfigure{\includegraphics[width=80mm]{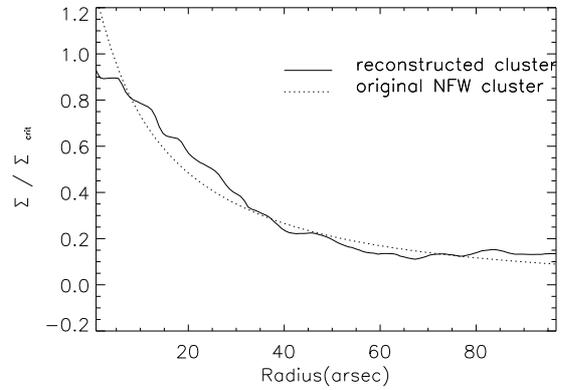}}
  \caption{Mass reconstruction obtained with the BGA and no overfitting  $\epsilon = 2 \times 10^{-10}$.
    {\it Top panel}: mass map after smoothing with a Gaussian. The mass inside the FOV is $M_{3.3'} = 6.1 \times 10^{14} M_\odot$, while the mass inside the core radius of 30'' is $M_{30''} = 1.39 \times 10^{14} M_\odot$.
    {\it Bottom panel}: Surface mass density profile (in units of $\Sigma_{crit}$) as a function of radius. Darker areas correspond to higher masses.} 
  \label{lens_case_301}
\end{figure} 
\begin{figure}
  \centering
  \subfigure{\includegraphics[width=80mm]{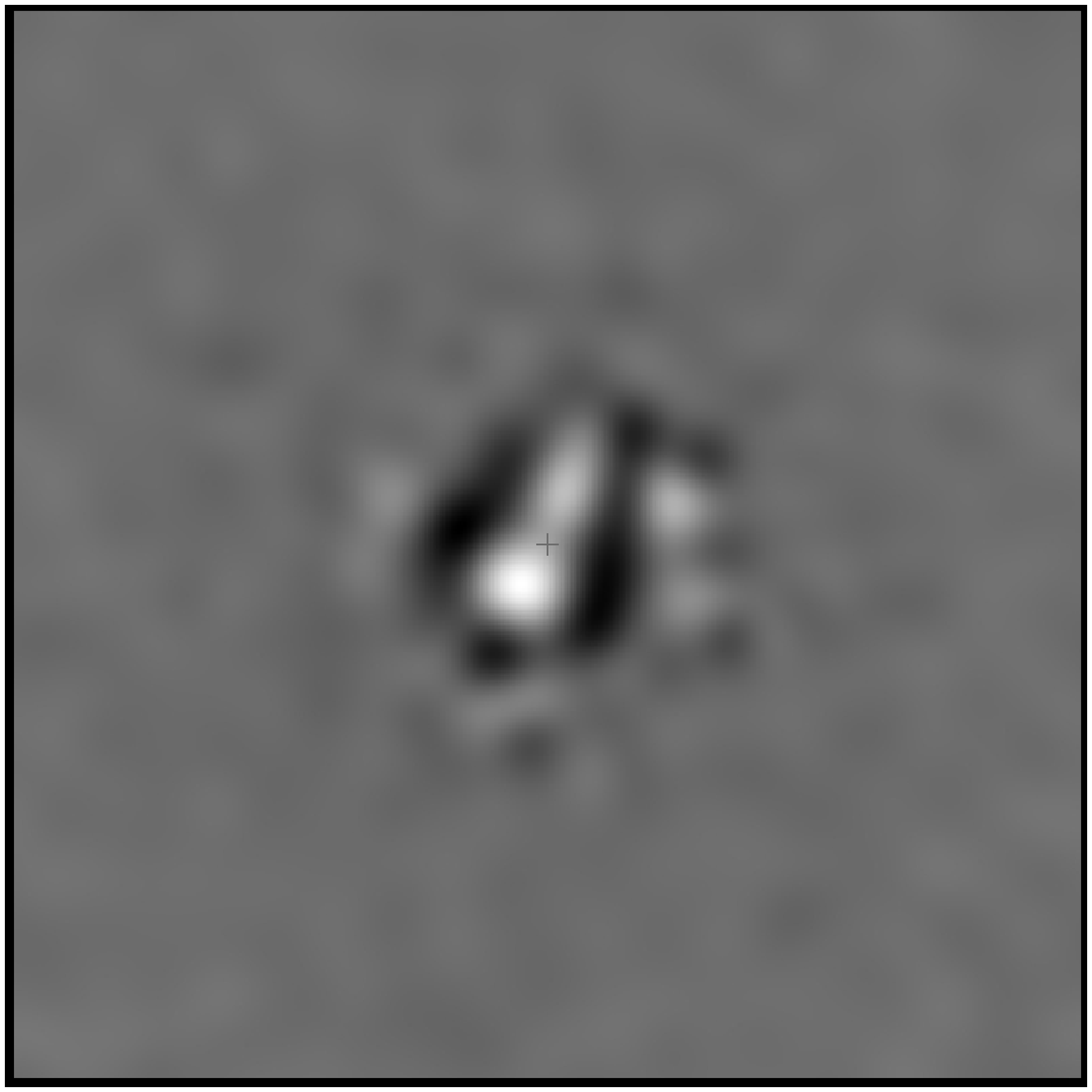}}
  \subfigure{\includegraphics[width=80mm]{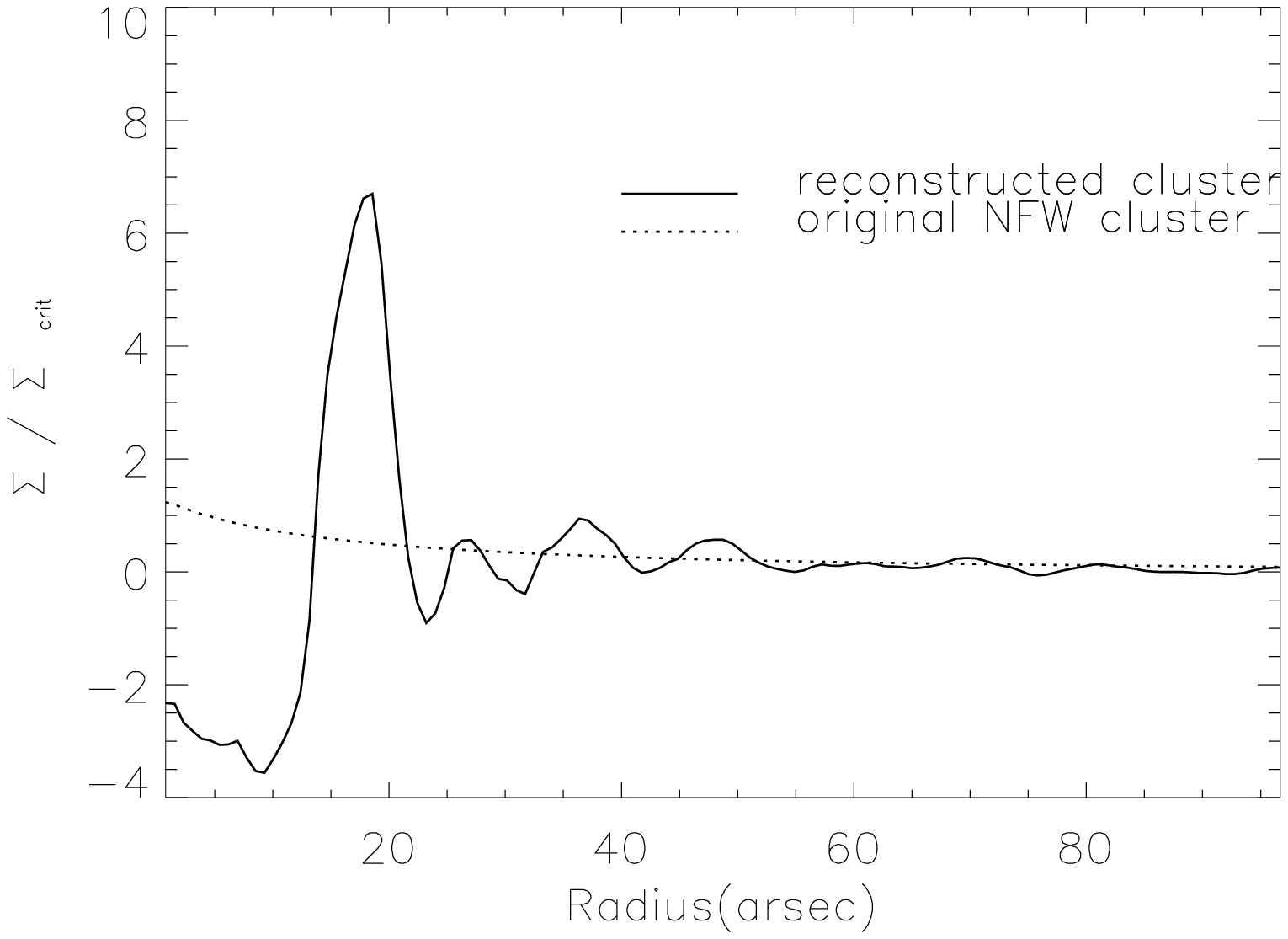}}
  \caption{Plots for $M_{3.3'} = 4.34 \times 10^{14} \msun$ and $M_{30''} \approx 1.9 \times 10^{14} \msun$. Overfitting case. It shows the solution obtained with the BGA when the method is forced to find a nearly exact solution to the problem ($\epsilon =2\times 10^{-15}$). The density profile inside the core radius does not follow the profile of the input NFW cluster. Different density peaks and dips can be seen around the center of the FOV. Darker areas correspond to higher masses.} 
  \label{lens_case_302}
\end{figure}
\begin{figure}
  \centering
  \includegraphics[width=80mm]{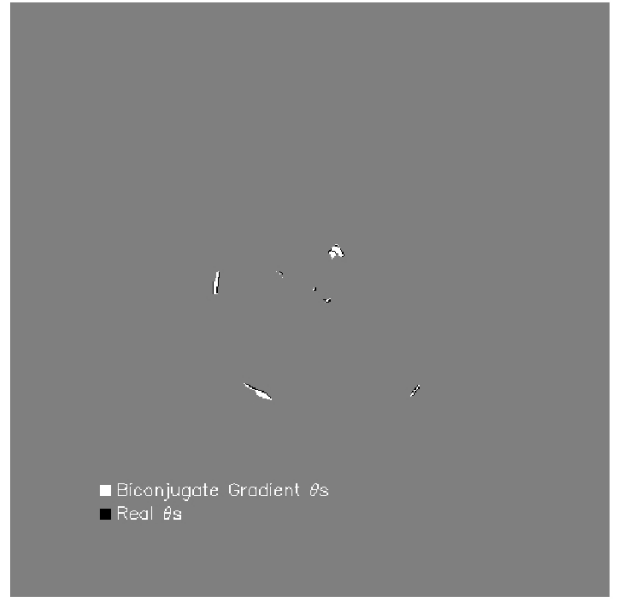}
  \caption{Black color indicates the observed (or true) arcs and in white we show the predicted arcs obtained with the solution shown in Fig. \ref{lens_case_302}.}
  \label{arcs1}
\end{figure}
In Fig.~\ref{arcs2}, we show the true strong lensing or $\theta$-map used to reconstruct the mass, compared with the predicted one derived from the optimal solution $X$. The black arcs are obtained from the equation $\theta = \Gamma M + \beta$, where the matrix $\Gamma$ is built from the real $\theta$ positions and the $32\times32$ cells, the vector $\beta$ contains the real positions of the background sources, and the vector $M$ contains the mean mass sampled in the $32\times32$ cells. 

The first interesting conclusion we can derive from this exercise is that the arcs predicted from the optimal solution differ significantly from the true observed arcs. This is unsurprising as the optimal solution lacks the resolution of the true underlying mass and hence we should expect a different set of strong lensed arcs. To reproduce the observed arcs, the solution has to {\it bend} the light in a different way. This can only be achieved with a mass distribution that is different (i.e biased away) from the true one.\\ 

This exercise summarizes the entire philosophy behind this paper: {\it using a non-parametric method with a uniform cell size, it is impossible to predict correctly the strong lensing data with an unbiased solution of the true underlying mass}. By default, the non-parametric method makes the {\it incorrect} assumption that the mass distribution is discrete and ignores the details of the mass distribution on scales smaller than the cell size. Hence, the derived solution has to be biased by the method in order to fit the data and compensate for this incorrect assumption. The best we can hope for is a solution that resembles the true underlying mass distribution but is unable to fit the observed data perfectly. This margin of error in the description of the observed data will then compensate the original error made by assuming that the mass is discretized. 
However, we note that we seek a solution as close as possible to the true solution, which can only be achieved when a realistic error, $R$, is allowed in the minimization of the system of linear equations given in Eq. (\ref{residual}). 

\section{Mass reconstruction} \label{sectVI}

To solve Eq. (\ref{main_system}), the lens plane is divided into a regular grid of $32\times32$ cells. This number is smaller than the number of constraints provided by the weak and strong lensing data. The mass in each cell plus the positions of the background strong lensing galaxies form a vector of unknown variables $X$ that has 1030 elements (1024 for the  mass cells and 6 for the three sources, each one with the $x$ and $y$ coordinates of the position of the background galaxy).\\ 

\subsection{The bi-conjugate gradient algorithm solution}
The bi-conjugate gradient algorithm (BGA) is a fast and powerful algorithm for finding the solutions of a system of linear equations. As mentioned earlier, rather than finding the exact solution, we seek an approximated one with an error large enough to compensate for the discretized mass and that the background galaxies are not point-like. The minimization is stopped at a point where $R^2 \approx \epsilon$. The choice of $\epsilon$ is based on the physical size of the background galaxies and also that the optimal solution should not reconstruct the data  perfectly as discussed in the previous subsection. A value of $\epsilon$ can be computed from the equation
\begin{equation}
\epsilon = \sum_i^{N_\theta}r_{\rm k}^2,
\label{calc_eps}
\end{equation}
where $r_{\rm k} = \Gamma^{\rm T} R_{\rm SL,prior} + \Gamma^{\rm T} R_{\rm WL,prior}$ contains an estimate of the physical size of the background galaxies ($R_{\rm SL,prior}$) and the error in the weak lensing measurements ($R_{\rm WL,prior}$, see previous sections for the definition of $\epsilon$ and its relation to $r_{\rm k}$). 

Once the value of $\epsilon$ is estimated, we can solve for the mass and position of the background sources. In Fig.~\ref{lens_case_301}, we show the mass reconstruction obtained with the BGA for a value of $\epsilon = 1\times 10^{-10}$ (computed in Eq. \ref{calc_eps}, corresponding to a $\sigma_{\rm SL} \sim 1.2$ arcsec and $\sigma_{\rm WL} = 0.3 $ or 30\%). The total recovered mass inside the FOV is $M(<3.3') = 6.1 \times 10^{14} M_\odot$, while $M (r<30'') = 1.39 \times 10^{14}M_\odot$. The radial density profile is shown in the bottom panel of the figure, where it is compared with the true mass profile.\\
Values of $\epsilon$ significantly smaller than $\sim 10^{-10}$ would produce an {\it overfitting} of the data, introducing systematics in the final mass reconstruction. A typical case of overfitting is shown in Fig.~\ref{lens_case_302}, where the threshold value of $\epsilon$ has been lowered several orders of magnitude ($\epsilon = 2 \times 10^{-15}$). This value pushes the solution to the limit of the BGA and allows us to predict almost perfectly the observed data. However, this solution is clearly biased with respect to the true underlying mass as is clear when looking at the density profile (bottom panel).
\begin{figure}
\centering
\includegraphics[width=80mm]{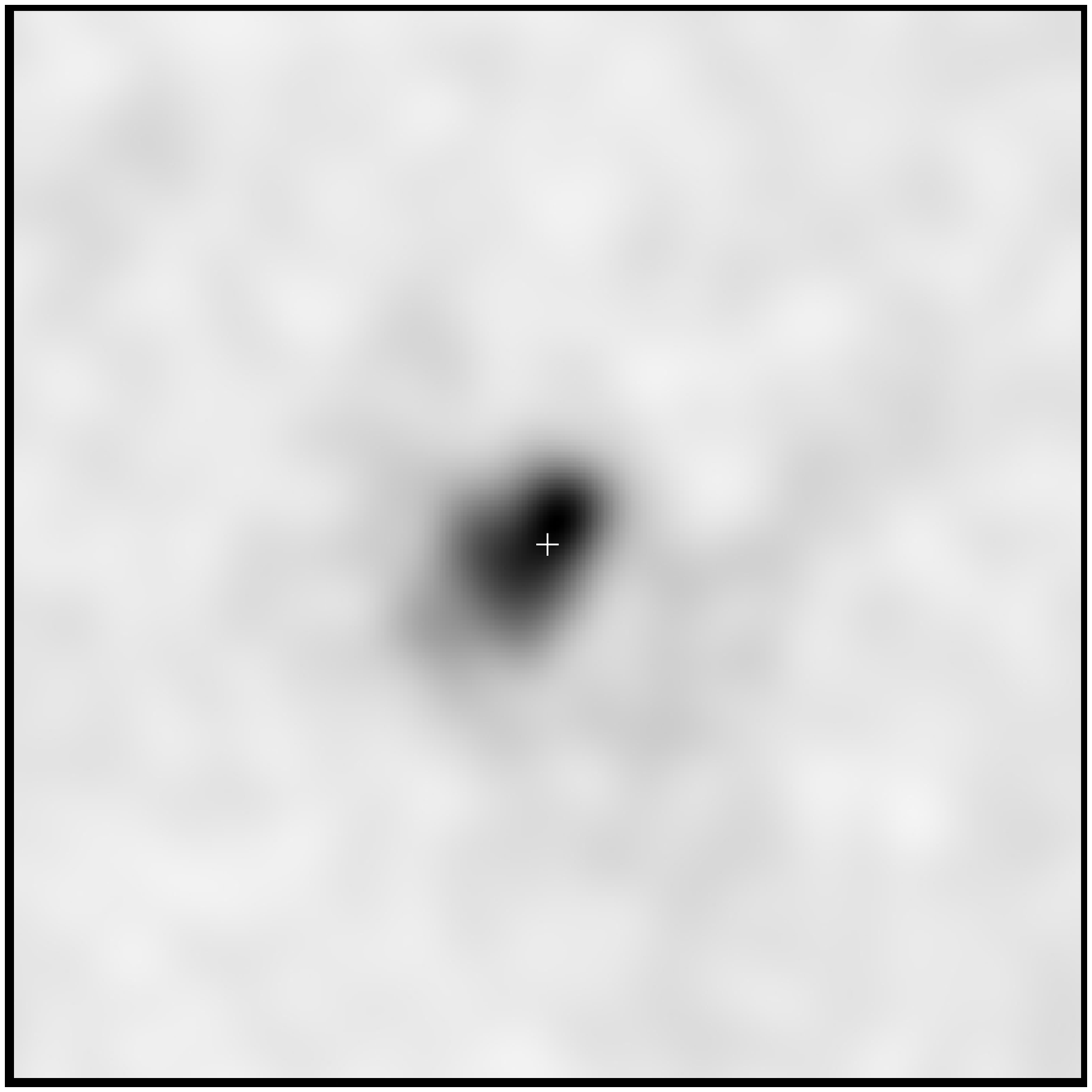}
\includegraphics[width=80mm]{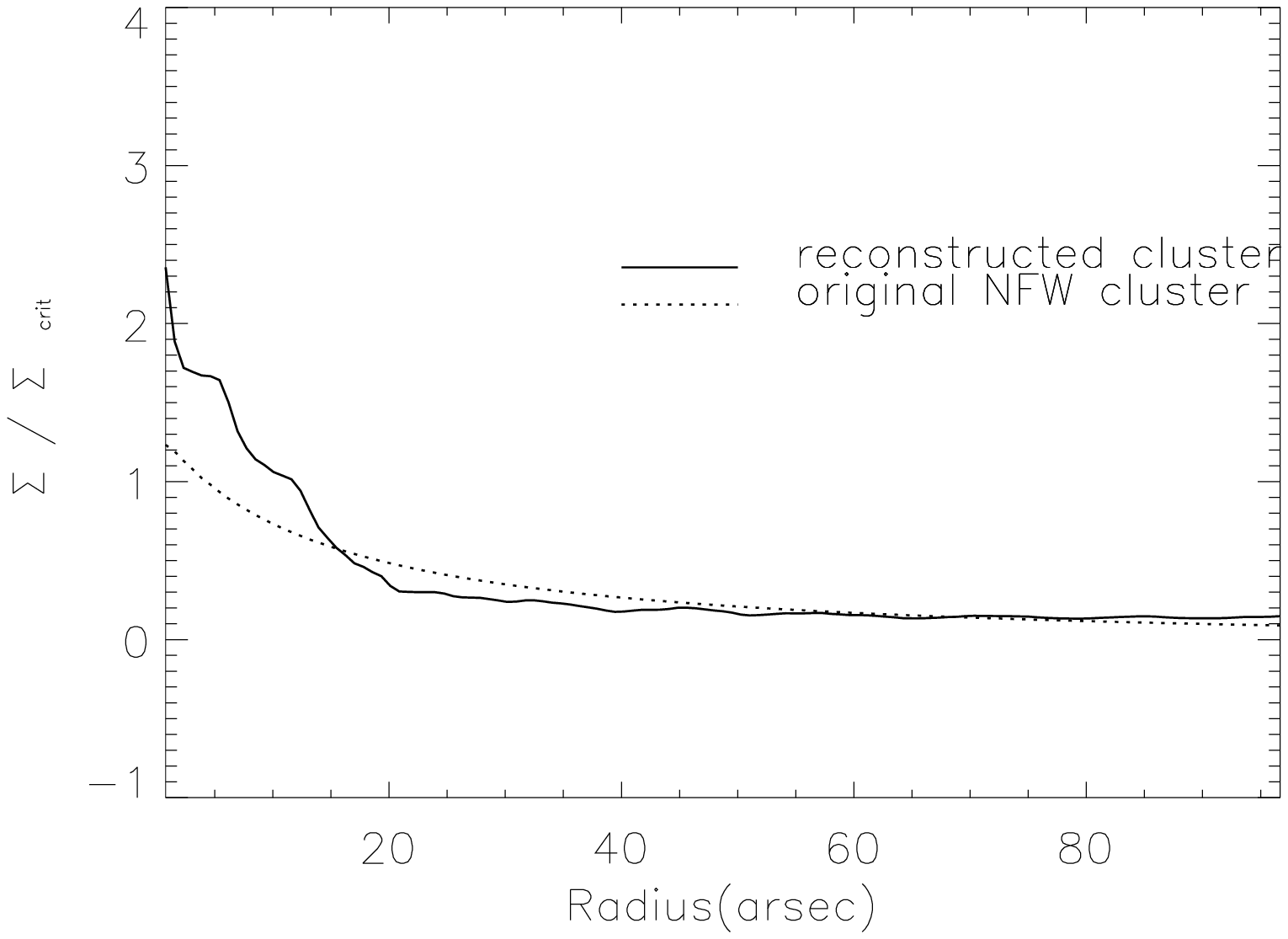}
\caption{Plots for $M_{3.3'} = 5.92 \times 10^{14} \msun$ and $M_{30''} = 1.22 \times 10^{14} \msun $. Mass reconstruction obtained with the QADP after 100 iterations. This case corresponds to a reasonable value of $\epsilon$ and can be compared with the BGA solution shown in Fig.~\ref{lens_case_301}. 
The QADP recovers a higher mass in the central region ($\Sigma/\Sigma_{crit}$) than the BGA.} 
\label{lens_case_303}
\end{figure}

The mass map shown in Fig.~\ref{lens_case_302} is obviously a poor solution in the sense that it deviates significantly from the underlying mass distribution. However, from the point of view of the system of linear equations it is a good solution because it is able to reproduce the data accurately. This is shown in  Fig.~\ref{arcs1}, where the observed arcs are compared to the predicted ones by the overfitting solution. 
This result should be compared with the case in Fig.~\ref{arcs2} showing the opposite situation where the closest representation of the mass distribution leads to an error in the predicted strongly lensed arcs. The conclusion we can extract from this example is that a simultaneous ({\it unbiased}) reconstruction of the mass and the lensing data is impossible with a non-parametric method that lacks the details of the mass distribution. 

\subsection{The quadratic programing algorithm solution}
The solution $X$ derived from the BGA might predict negative masses, which could lead to large fluctuations in the mass density profile as the negative fluctuations have to be compensated for by larger positive fluctuations. However, \citet[][and references within]{Hoekstra_et_al_2011} report that {\it cosmic noise} (an induced shear effect by uncorrelated halos and large-scale structure) has to be taken into account when estimating the error bars in any cluster mass reconstruction that might lead to a negative convergence in the regime of the weak lensing. So a negative convergence is not completely unrealistic. 

To avoid the large fluctuations at small radii exhibited by the biconjugate gradient, which can indicate a non-physical solution, we use the quadratic programming algorithm (QADP, see Appendix B), which prevents negative masses from appearing in the solution. This method resembles the maximum entropy method introduced in \cite{Jee_et_al_2007}, since both impose a positive prior on the mass.

The QADP has a smooth behavior in the inner regions, where no large fluctuations are found, even in the crucial areas of the lens plane where the transition between the WL and SL regimes is observed. In addition, QADP provides an independent solution that should agree with the one derived by the BGA.\\    
 
The number of iterations of the algorithm can be directly related to $\epsilon$. The overfitting solution obtained by the QADP algorithm converges only after a large number of iterations ($\sim 10^4 - 10^5$) or equivalently after defining a small value for $\epsilon$. \\

In Fig.~\ref{lens_case_303}, we show the solution obtained with QADP after 100 iterations. This result can be compared with the one in Fig.~\ref{lens_case_301}. The QADP recovers a higher mass than the BGA in the central region. 

\begin{figure}
  \centering
\subfigure{\includegraphics[width=80mm]{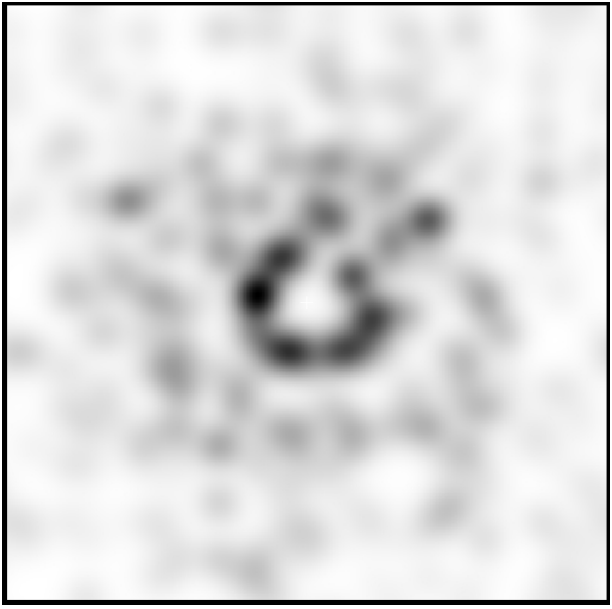}}
\subfigure{\includegraphics[width=80mm]{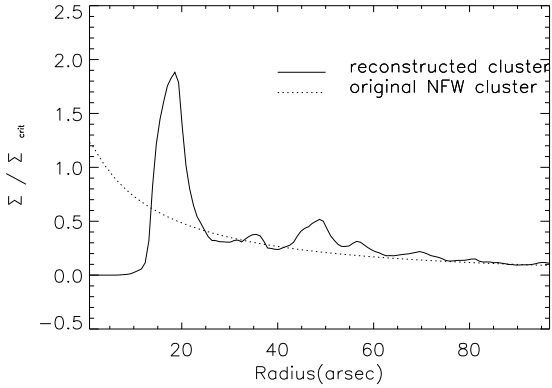}}
    \caption{$M_{3.3'} = 6.81 \times 10^{14} \msun $ and $M_{30''} = 1.59 \times 10^{14}$. Mass reconstruction obtained with QADP and after $10^5$ iterations (overfitting case). The density peaks at $r \sim 15''$ and a bump is observed at $r \sim 50''$. Darker areas correspond to higher masses. }
    \label{lens_case_304}
\end{figure}

In Fig.~\ref{lens_case_304}, we show the overfitting case obtained with QADP with a large number of iterations ($N_{\rm iter}=10^5$, or similarly, with a very small value for $\epsilon$).~In this case, the mass is pushed away from the center towards larger radii in a similar way to what was observed using the BGA. This is more clearly evident in the density profile. A peak in the density is observed at $r=20''$ and an additional bump at $r=50''$. The way in which the WL and SL are weighted is different in both methods. The overfitting solution differs significantly from the true mass (and also from Jee's reconstruction in the central part). The overfitting solution is dominated in our case by the WL part of the data \citep[as in][]{Jee_et_al_2007}. As shown in Fig.~\ref{WL_overfit}, the WL alone case shows a mass deficit at the center that is compensated for by the ring in the outer regions. Whether a similar situation occurs in \cite{Jee_et_al_2007} is unclear but we note that Jee's mass reconstruction predicts a lower mass at the center than that of \cite{Zitrin_et_al_2009} \citep[as seen in figure 21 of][]{Umetsu_et_al_2010}.\\ 

In \citet{Jee_et_al_2007}, their Fig. 10 shows the radial mass density profile of the cluster, with a $\Sigma_{\rm c}$ given at a fiducial redshift of $z_{\rm f}=3$. The authors state that the resulting profile does not match any conventional analytic profile. The density, peaking at the center with the value of $\Sigma/\Sigma_{\rm c}=1.3$, rapidly decreases from the center to the end of the core radius at $r = 50''$. The profile then remains {\it almost} constant around a value of $\Sigma/\Sigma_{\rm c}=0.7$. Only at radius $r=70''$ from the center is an increment observable, extending out to $r = 80''$ with a peak at $r=75''$.  This is what the authors refer to as the {\it bump}. In two dimensions, this bump appears like a ring structure, separated from the core by 20''. 

A plateau was detected by \cite{Jee_et_al_2007} at  $r \sim 50''$, that was not found by \cite{Umetsu_et_al_2010}, who instead measured a monotonically decreasing density. This plateau might depend on the initial guess. The WL part of the data displays this plateau more than the SL data, especially in those regions where WL constraints are weaker. The role that the prior plays in determining the regularization term in the MEM has to be investigated in more detail and leaves questions open on how the choice of the prior could affect the radii outside the central core.  

\section{Discussion and conclusions} \label{sectVII}

The interesting analysis of \citet{Jee_et_al_2007} appears to detect a dark matter ring around the core of CL0024. This ring might have been caused by a recent high speed collision between two massive clusters along the line of sight. If confirmed, CL0024 would be an interesting laboratory to test different physical phenomena. We have explored the possibility that spurious ring-like structures might appear as a consequence of overfitting lensing data in a non-parametric way. We show how the optimal (unbiased) solution should produce a fit to the data significantly poorer than the minimal $\chi^2$ solution. This error is necessary to account for the initial error introduced when neglecting the impact of the small-scale fluctuations on the mass distribution. We demonstrate our argument by using a simulated data set  where all the variables are known a priori and the reconstructed mass can be compared with the original one. The simulation shows how overfitting the data introduces artifacts in the reconstructed solution, which can resemble the ring-like structure found in \citet{Jee_et_al_2007}. 
The methods in \citet{Jee_et_al_2007} and the one used in this work are different in some aspects but both methods share many common key features such as the lens plane is divided into a regular grid and the parameters to be constrained are basically those for convergence in the pixels. Hence both methods should also have the same systematic effects and in particular be sensitive in a similar way to overfitting. 

\begin{figure}
  \centering
  \includegraphics[width=80mm]{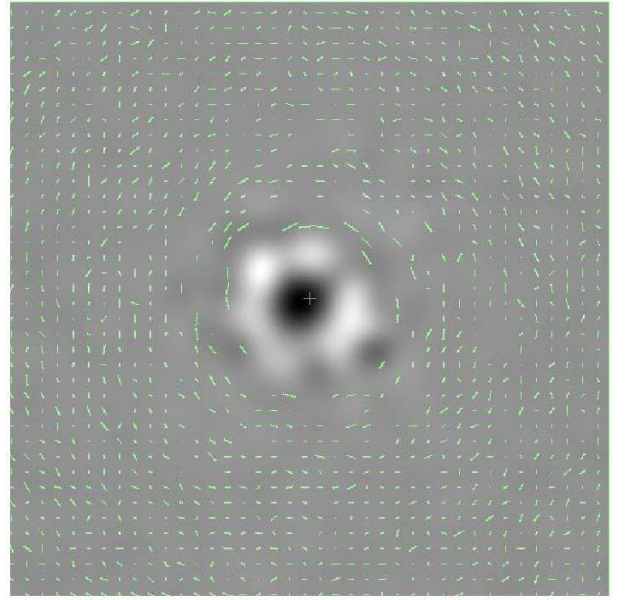}
  \caption{Reconstructed image for the case where only weak lensing data is used in the reconstruction. 
    A clear ring of matter appears in the area where the density of weak lensing data gets reduced. Whiter colors indicate more mass.}
  \label{WL_overfit}
\end{figure}

Another interesting feature shown by the simulations that needs to be investigated more (with the actual data) is that when the density of weak lensing data is non-uniform across the field of view, there is a tendency for the overfitted solution to increase the mass density in the areas with fewer weak lensing data. We show one example in Fig.~\ref{WL_overfit}, where only the simulated weak lensing data is used to find the solution. The plot shows the WL data overlaid on the overfitted solution found for this case. In the case of CL0024, we expect a lower density of WL points toward the center of the cluster owing to contamination by the cluster members. While the SL data constraints the inner central region of the cluster, the outer regions are basically constrained by the WL data alone. In-between these two regions, the density of WL data points should show a gradient, and the effect of the non-uniformity of the WL data points might have a negative effect on the solution. The reality of the ringlike structure will need to be investigated in more detail. 

We note that the covariance matrix of the residual might not necessarily be diagonal. As discussed in section 7 of \citet{Diego_et_al_2007}, the elements of the residual are correlated with each other, in particular the strong lensing part of the residual. The elements of the WL portion of the residual are more weakly correlated with each other, and the diagonal approximation is in this case more valid. This is particularly true in  our case where the error assigned to the WL measurements is the predominant one (30\%). 
Since the WL data are more relevant to understanding the ring-like structure, we adopt the diagonal approximation for the covariance matrix. In addition, the second reason why we prefer to adopt this approximation in this paper is that \citet{Jee_et_al_2007} assumed that the data are uncorrelated (the covariance matrix is diagonal for an uncorrelated residual). The issue of the effect of the covariance matrix in lensing reconstruction has not been addressed by any method (to the best of our knowledge) and we plan to do so in a future paper. 
Another interesting point that deserves discussion is that in \citet{Jee_et_al_2007} a regularization term is included in the analysis, among other things, to prevent overfitting. This regularization term, however, does not guarantee that overfitting is prevented. The main objective of the regularization term is to favor solutions that are smooth by introducing a prior that represents a smoothed version of the solution. If we consider the extreme case where the reconstructed solution converges to the prior in their regularization term (this is not an unrealistic scenario because the prior is updated at each iteration and based on the previous solution), the regularization term tends to zero forcing the other terms in $\chi^2$ to be even smaller and hence closer to an overfitting situation. The SL and WL terms to be minimized are the ones that really constrain the model and can still be too small even for smooth solutions. Our work shows that a good solution obtained with our non-parametric method should predict arcs significantly different from the ones observed. Only when overfitting is allowed can the reconstructed data closely reproduce the observations (see Figs.~\ref{arcs2} and \ref{arcs1} above).

Our work shows the validity and usefulness of non-parametric methods but also shows some of its limitations, in particular that one should not be too ambitious when fitting the data.

\begin{acknowledgements}
We acknowledge partial financial support from the Ministerio de Ciencia e Innovaci\'on project AYA2007-68058-C03-02.~PPP acknowledges support from the Spanish Ministerio de Educaci\'on y Ciencia and CSIC for an I3P grant.
\end{acknowledgements}

\bibliographystyle{aa}
\nocite{*}
\bibliography{lensing_bib}

\begin{thebibliography}{45}
\expandafter\ifx\csname natexlab\endcsname\relax\def\natexlab#1{#1}\fi

\bibitem[{{Abdelsalam} {et~al.}(1998{\natexlab{a}}){Abdelsalam}, {Saha}, \&
  {Williams}}]{Abdelsalam_et_al_1998}
{Abdelsalam}, H.~M., {Saha}, P., \& {Williams}, L.~L.~R. 1998{\natexlab{a}},
  \aj, 116, 1541

\bibitem[{{Abdelsalam} {et~al.}(1998{\natexlab{b}}){Abdelsalam}, {Saha}, \&
  {Williams}}]{Abdelsalam_et_al_1998b}
{Abdelsalam}, H.~M., {Saha}, P., \& {Williams}, L.~L.~R. 1998{\natexlab{b}},
  \aj, 116, 1541

\bibitem[{{Brada{\v c}} {et~al.}(2005){Brada{\v c}}, {Schneider}, {Lombardi},
  \& {Erben}}]{Bradac_et_al_2005}
{Brada{\v c}}, M., {Schneider}, P., {Lombardi}, M., \& {Erben}, T. 2005, \aap,
  437, 39

\bibitem[{{Bridle} {et~al.}(1998){Bridle}, {Hobson}, {Lasenby}, \&
  {Saunders}}]{Bridle_et_al_1998}
{Bridle}, S.~L., {Hobson}, M.~P., {Lasenby}, A.~N., \& {Saunders}, R. 1998,
  \mnras, 299, 895

\bibitem[{{Broadhurst} {et~al.}(2005{\natexlab{a}}){Broadhurst},
  {Ben{\'{\i}}tez}, {Coe}, {Sharon}, {Zekser}, {White}, {Ford}, {Bouwens},
  {Blakeslee}, {Clampin}, {Cross}, {Franx}, {Frye}, {Hartig}, {Illingworth},
  {Infante}, {Menanteau}, {Meurer}, {Postman}, {Ardila}, {Bartko}, {Brown},
  {Burrows}, {Cheng}, {Feldman}, {Golimowski}, {Goto}, {Gronwall}, {Herranz},
  {Holden}, {Homeier}, {Krist}, {Lesser}, {Martel}, {Miley}, {Rosati},
  {Sirianni}, {Sparks}, {Steindling}, {Tran}, {Tsvetanov}, \&
  {Zheng}}]{Broadhurst_et_al_2005B}
{Broadhurst}, T., {Ben{\'{\i}}tez}, N., {Coe}, D., {et~al.} 2005{\natexlab{a}},
  \apj, 621, 53

\bibitem[{{Broadhurst} {et~al.}(2000){Broadhurst}, {Huang}, {Frye}, \&
  {Ellis}}]{Broadhurst_et_al_2000}
{Broadhurst}, T., {Huang}, X., {Frye}, B., \& {Ellis}, R. 2000, \apjl, 534, L15

\bibitem[{{Broadhurst} {et~al.}(2005{\natexlab{b}}){Broadhurst}, {Takada},
  {Umetsu}, {Kong}, {Arimoto}, {Chiba}, \& {Futamase}}]{Broadhurst_et_al_2005A}
{Broadhurst}, T., {Takada}, M., {Umetsu}, K., {et~al.} 2005{\natexlab{b}},
  \apjl, 619, L143

\bibitem[{{Cacciato} {et~al.}(2006){Cacciato}, {Bartelmann}, {Meneghetti}, \&
  {Moscardini}}]{Cacciato_et_al_2006}
{Cacciato}, M., {Bartelmann}, M., {Meneghetti}, M., \& {Moscardini}, L. 2006,
  \aap, 458, 349

\bibitem[{{Clowe} {et~al.}(2006){Clowe}, {Brada{\v c}}, {Gonzalez},
  {Markevitch}, {Randall}, {Jones}, \& {Zaritsky}}]{bullet_cluster}
{Clowe}, D., {Brada{\v c}}, M., {Gonzalez}, A.~H., {et~al.} 2006, \apjl, 648,
  L109

\bibitem[{{Colley} {et~al.}(1996){Colley}, {Tyson}, \&
  {Turner}}]{Colley_et_al_1996}
{Colley}, W.~N., {Tyson}, J.~A., \& {Turner}, E.~L. 1996, \apjl, 461, L83+

\bibitem[{{Comerford} {et~al.}(2006){Comerford}, {Meneghetti}, {Bartelmann}, \&
  {Schirmer}}]{Comerford_et_al_2006}
{Comerford}, J.~M., {Meneghetti}, M., {Bartelmann}, M., \& {Schirmer}, M. 2006,
  \apj, 642, 39

\bibitem[{{Czoske} {et~al.}(2001){Czoske}, {Kneib}, {Soucail}, {Bridges},
  {Mellier}, \& {Cuillandre}}]{Czoske_et_al_2001}
{Czoske}, O., {Kneib}, J., {Soucail}, G., {et~al.} 2001, \aap, 372, 391

\bibitem[{{Diego} {et~al.}(2005{\natexlab{a}}){Diego}, {Protopapas}, {Sandvik},
  \& {Tegmark}}]{Diego_et_al_2005a}
{Diego}, J.~M., {Protopapas}, P., {Sandvik}, H.~B., \& {Tegmark}, M.
  2005{\natexlab{a}}, \mnras, 360, 477

\bibitem[{{Diego} {et~al.}(2005{\natexlab{b}}){Diego}, {Sandvik}, {Protopapas},
  {Tegmark}, {Ben{\'{\i}}tez}, \& {Broadhurst}}]{Diego_et_al_2005b}
{Diego}, J.~M., {Sandvik}, H.~B., {Protopapas}, P., {et~al.}
  2005{\natexlab{b}}, \mnras, 362, 1247

\bibitem[{{Diego} {et~al.}(2007){Diego}, {Tegmark}, {Protopapas}, \&
  {Sandvik}}]{Diego_et_al_2007}
{Diego}, J.~M., {Tegmark}, M., {Protopapas}, P., \& {Sandvik}, H.~B. 2007,
  \mnras, 375, 958

\bibitem[{{Dye} {et~al.}(2001){Dye}, {Taylor}, {Thommes}, {Meisenheimer},
  {Wolf}, \& {Peacock}}]{Dye_et_al_2001}
{Dye}, S., {Taylor}, A.~N., {Thommes}, E.~M., {et~al.} 2001, \mnras, 321, 685

\bibitem[{{Halkola} {et~al.}(2006){Halkola}, {Seitz}, \&
  {Pannella}}]{Halkola_et_al_2006}
{Halkola}, A., {Seitz}, S., \& {Pannella}, M. 2006, \mnras, 372, 1425

\bibitem[{{Hoekstra} {et~al.}(2011){Hoekstra}, {Hartlap}, {Hilbert}, \& {van
  Uitert}}]{Hoekstra_et_al_2011}
{Hoekstra}, H., {Hartlap}, J., {Hilbert}, S., \& {van Uitert}, E. 2011, \mnras,
  412, 2095

\bibitem[{{Jee} {et~al.}(2007){Jee}, {Ford}, {Illingworth}, {White},
  {Broadhurst}, {Coe}, {Meurer}, {van der Wel}, {Ben{\'{\i}}tez}, {Blakeslee},
  {Bouwens}, {Bradley}, {Demarco}, {Homeier}, {Martel}, \&
  {Mei}}]{Jee_et_al_2007}
{Jee}, M.~J., {Ford}, H.~C., {Illingworth}, G.~D., {et~al.} 2007, \apj, 661,
  728

\bibitem[{{Kaiser} \& {Squires}(1993)}]{kaiser_squires_1993}
{Kaiser}, N. \& {Squires}, G. 1993, \apj, 404, 441

\bibitem[{{Kneib} {et~al.}(2003){Kneib}, {Hudelot}, {Ellis}, {Treu}, {Smith},
  {Marshall}, {Czoske}, {Smail}, \& {Natarajan}}]{Kneib_et_al_2003}
{Kneib}, J., {Hudelot}, P., {Ellis}, R.~S., {et~al.} 2003, \apj, 598, 804

\bibitem[{{Milgrom} \& {Sanders}(2008)}]{Milgrom_Sanders_2008}
{Milgrom}, M. \& {Sanders}, R.~H. 2008, \apj, 678, 131

\bibitem[{{Navarro} {et~al.}(1996){Navarro}, {Frenk}, \& {White}}]{NFW}
{Navarro}, J.~F., {Frenk}, C.~S., \& {White}, S.~D.~M. 1996, \apj, 462, 563

\bibitem[{{Ota} {et~al.}(2004){Ota}, {Pointecouteau}, {Hattori}, \&
  {Mitsuda}}]{Ota_et_al_2004}
{Ota}, N., {Pointecouteau}, E., {Hattori}, M., \& {Mitsuda}, K. 2004, \apj,
  601, 120

\bibitem[{{Padmanabhan}(2002)}]{Padmanabhan_vol_3}
{Padmanabhan}, T. 2002, {Theoretical Astrophysics - Volume 3, Galaxies and
  Cosmology}, ed. {Padmanabhan, T.}

\bibitem[{{Press} {et~al.}(1997){Press}, {Teukolsky}, {Vetterling}, \&
  {Flannery}}]{numerical_recipes}
{Press}, W.~H., {Teukolsky}, S.~A., {Vetterling}, W.~T., \& {Flannery}, B.~P.
  1997, {Numerical Recipes in Fortran 77}

\bibitem[{{Qin} {et~al.}(2008){Qin}, {Shan}, \& {Tilquin}}]{Qin_et_al_2008}
{Qin}, B., {Shan}, H., \& {Tilquin}, A. 2008, \apjl, 679, L81

\bibitem[{{Saha} \& {Williams}(1997)}]{Saha_Williams_1997}
{Saha}, P. \& {Williams}, L.~L.~R. 1997, \mnras, 292, 148

\bibitem[{{Saha} {et~al.}(1999){Saha}, {Williams}, \&
  {AbdelSalam}}]{Saha_Williams_Abdelsalam_1999}
{Saha}, P., {Williams}, L.~L.~R., \& {AbdelSalam}, H. 1999, ArXiv:
  astro-ph/9909249

\bibitem[{{Schneider} \& {Seitz}(1995)}]{Schneider_Seitz_1995}
{Schneider}, P. \& {Seitz}, C. 1995, \aap, 294, 411

\bibitem[{{Seitz} {et~al.}(1998){Seitz}, {Schneider}, \&
  {Bartelmann}}]{seitz_et_al_1998}
{Seitz}, S., {Schneider}, P., \& {Bartelmann}, M. 1998, \aap, 337, 325

\bibitem[{Sha {et~al.}(2002)Sha, Saul, \& Lee}]{Sha_Saul_Lee}
Sha, F., Saul, L.~K., \& Lee, D.~D. 2002, in Advances in Neural Information
  Processing Systems 15 (MIT Press), 1041--1048

\bibitem[{{Smail} {et~al.}(1996){Smail}, {Dressler}, {Kneib}, {Ellis}, {Couch},
  {Sharples}, \& {Oemler}}]{Smail_et_al_1996}
{Smail}, I., {Dressler}, A., {Kneib}, J., {et~al.} 1996, \apj, 469, 508

\bibitem[{{Smith} {et~al.}(2005){Smith}, {Kneib}, {Smail}, {Mazzotta},
  {Ebeling}, \& {Czoske}}]{Smith_et_al_2005}
{Smith}, G.~P., {Kneib}, J., {Smail}, I., {et~al.} 2005, \mnras, 359, 417

\bibitem[{{Taylor} {et~al.}(1998){Taylor}, {Dye}, {Broadhurst}, {Benitez}, \&
  {van Kampen}}]{Taylor_et_al_1998}
{Taylor}, A.~N., {Dye}, S., {Broadhurst}, T.~J., {Benitez}, N., \& {van
  Kampen}, E. 1998, \apj, 501, 539

\bibitem[{{Tyson} \& {Fischer}(1995)}]{Tyson_Fischer_1995}
{Tyson}, J.~A. \& {Fischer}, P. 1995, \apjl, 446, L55+

\bibitem[{{Tyson} {et~al.}(1998){Tyson}, {Kochanski}, \&
  {dell'Antonio}}]{Tyson_et_al_1998}
{Tyson}, J.~A., {Kochanski}, G.~P., \& {dell'Antonio}, I.~P. 1998, \apjl, 498,
  L107+

\bibitem[{{Tyson} {et~al.}(1990){Tyson}, {Wenk}, \&
  {Valdes}}]{Tyson_et_al_1990}
{Tyson}, J.~A., {Wenk}, R.~A., \& {Valdes}, F. 1990, \apjl, 349, L1

\bibitem[{{Umetsu} {et~al.}(2011){Umetsu}, {Broadhurst}, {Zitrin},
  {Medezinski}, \& {Hsu}}]{Umetsu_et_al_2011}
{Umetsu}, K., {Broadhurst}, T., {Zitrin}, A., {Medezinski}, E., \& {Hsu}, L.-Y.
  2011, \apj, 729, 127

\bibitem[{{Umetsu} {et~al.}(2010){Umetsu}, {Medezinski}, {Broadhurst},
  {Zitrin}, {Okabe}, {Hsieh}, \& {Molnar}}]{Umetsu_et_al_2010}
{Umetsu}, K., {Medezinski}, E., {Broadhurst}, T., {et~al.} 2010, \apj, 714,
  1470

\bibitem[{{Wallington} {et~al.}(1992){Wallington}, {Kochanek}, \&
  {Narayan}}]{Wallington_et_al_1992}
{Wallington}, S., {Kochanek}, C.~S., \& {Narayan}, R. 1992, in Bulletin of the
  American Astronomical Society, Vol.~24, Bulletin of the American Astronomical
  Society, 1192

\bibitem[{{Zhang} {et~al.}(2005){Zhang}, {B{\"o}hringer}, {Mellier}, {Soucail},
  \& {Forman}}]{Zhang_et_al_2005}
{Zhang}, Y., {B{\"o}hringer}, H., {Mellier}, Y., {Soucail}, G., \& {Forman}, W.
  2005, \aap, 429, 85

\bibitem[{{Zitrin} {et~al.}(2009){Zitrin}, {Broadhurst}, {Umetsu}, {Coe},
  {Ben{\'{\i}}tez}, {Ascaso}, {Bradley}, {Ford}, {Jee}, {Medezinski},
  {Rephaeli}, \& {Zheng}}]{Zitrin_et_al_2009}
{Zitrin}, A., {Broadhurst}, T., {Umetsu}, K., {et~al.} 2009, \mnras, 396, 1985

\bibitem[{{Zu Hone} {et~al.}(2009){Zu Hone}, {Lamb}, \&
  {Ricker}}]{ZuHone_et_al_2009}
{Zu Hone}, J.~A., {Lamb}, D.~Q., \& {Ricker}, P.~M. 2009, \apj, 696, 694

\bibitem[{{Zwicky}(1959)}]{Zwicky_1959}
{Zwicky}, F. 1959, Handbuch der Physik, 53, 390

\end{thebibliography}

\appendix

\section{How is built the $\Gamma$ matrix}
The $\Gamma$ matrix is the basis of the Weak and Strong Lensing Analysis Package (WSLAP) and contains the information about how each cell in the grid contributes to either the $j^{\rm th}$ deflection angle or the $k^{\rm th}$ shear measurement. In the SL case, it also contains information about the source identity of the $j^{\rm th}$ pixel in a given lensed arc. All this information is organized in rows, each row corresponding to one constraint (deflection angle for SL and shear for WL). The final structure of $\Gamma$ is
\begin{equation}\label{gamma_matrix}
{\bf \Gamma} = \left |
\begin{array}{ccc}
{\bf \Upsilon}_{\rm \bf x} & \bf 1 & \bf 0\\
\bf \Upsilon_{\rm \bf y} & \bf 0 & \bf 1\\
\bf \Delta_1 & \bf 0 & \bf 0\\
\bf \Delta_2 & \bf 0 & \bf 0\\
\end{array} 
\right |.
\end{equation}
The specific form of the ${\bf \Upsilon}$ and {$\bf \Gamma$} matrices depends on the choice of basis system. For clarity purposes, we assume that this system is based on Gaussians positioned on the grid. This grid is a division of the lens plane into cells, where the mass in a cell is assumed to be distributed as a Gaussian of dispersion $\sigma$, which is proportional to the size of the cell. A proportionality factor $\sim 2$ gives very satisfactory results in terms of reproducing the constraints. The integrated mass at a given distance $\delta$ from the center of the cell is then
\begin{equation}
M(\delta) = 1 - \exp (\delta^2/2\sigma^2).
\end{equation}
Since the basis has circular symmetry, the $x$ and $y$ components of the deflection angle $\alpha$ at the same point can be estimated easily as
\begin{equation}\label{alpha_x}
\alpha_{\rm x} (\delta) = \Upsilon_{\rm x} = \lambda[1-\exp (-\delta^2/1\sigma^2)]\frac{\delta_{\rm x}}{\delta^2},
\end{equation}
\begin{equation}\label{alpha_y}
\alpha_{\rm y} (\delta) = \Upsilon_{\rm y} = \lambda[1-\exp (-\delta^2/1\sigma^2)]\frac{\delta_{\rm y}}{\delta^2},
\end{equation}
where the multiplying constant $\lambda$ contains all the cosmological and redshift dependence
\begin{equation}\label{lambda}
\lambda = 10^{15} M_\odot \frac{4G}{c^2} \frac{D_{\rm ls}}{D_{\rm ol} D_{\rm os}} h^{-1} \mbox{ rad }.
\end{equation}
The factor $\delta_{\rm x}$ in Eq. (\ref{alpha_x}) is just the difference (in radians) between the x position in the arc (x of pixel $\theta_{\rm x}$) and the x position of the cell $j$ in the grid ($\delta_{\rm x} = \theta_{\rm x}(i)-\theta'_{\rm x}(j)$). Similarly, we can define $\delta_{\rm y} = \theta_{\rm y}(i) -\theta'_{\rm y}(j)$ and $\delta = \sqrt{\delta_{\rm x}^2 + \delta_{\rm y}^2}$.

The $\mathbf {\Delta_1}$ and $\mathbf {\Delta_2}$ matrices can be computed in a similar way but in this case, since we need to calculate the derivatives, the deflection angles $\alpha_{\rm x}$ and $\alpha_{\rm y}$ have to be computed at three points $\delta_1$, $\delta_2$, and $\delta_3$. The first point, $\delta_1$, is the same as $\delta$ above. The second and third points ($\delta_2$ and $\delta_3$) are one (or a few) pixel(s) left (or right) and up (or down) the pixel at $\delta_1$, respectively. Then $\mathbf {\Delta}_1$ is just the difference 
\begin{equation}
{\mathbf \Delta}_1 = \frac{1}{2} \frac{[\alpha_{\rm x}(\delta_3)-\alpha_{\rm x}(\delta_1)]-[\alpha_{\rm y}(\delta_3)-\alpha_{\rm y}(\delta_1)]}{\mbox{pix2rad}},
\end{equation}
\begin{equation}
{\mathbf \Delta}_2 =\frac{\alpha_{\rm x}(\delta_3)-\alpha_{\rm x}(\delta_1)}{\mbox{pix2rad}}= \frac{\alpha_{\rm y}(\delta_3)-\alpha_{\rm y}(\delta_1)}{\mbox{pix2rad}},
\end{equation}
where pix2rad is the size of the pixel in radians. Since we included the factor $10^{15}M_{\odot}$ in $\lambda$ (see Eq. \ref{lambda}), the mass in the solution vector will be given in $10^{15} h^{-1}M_\odot$ units. The $h^{-1}$ dependency exists because in $\lambda$ we have the ratio $D_{\rm ls}/(D_{\rm ol}D_{\rm os})$, which goes as $h$.

The ${\mathbf 0}$ (null) and ${\mathbf 1}$ (0's and 1's) matrices on the right side of $\Gamma$ add $2N_{\rm s}$ additional columns. The bottom part of thess columns consist entirely of 0's since the shear measurements are independent of the position $\beta$ of the sources. The $N_\theta \times N_{\rm s}$ dimensional matrices ${\mathbf 1}$ contain 1's in the $ij$ positions ($i \in [1,N_\theta],j \in [1,N_{\rm s}]$), where the $i^{\rm th}$ $\theta$ pixel comes from the $j$ source and 0's elsewhere. 

\section{Minimizing algorithms}

\subsection{Biconjugate gradient algorithm or BGA}
The biconjugate gradient \citep{numerical_recipes} algorithm is one of the fastest and most powerful algorithms for solving systems of linear equations. It is also extremely useful for finding approximate solutions for systems where no exact solutions exists or where the exact solution is not the one we are interested in. The latter is our case. Given a system of linear equations 
\begin{equation}
Ax = b,
\label{linear_system}
\end{equation}
a solution of this system can be found by minimizing the function
\begin{equation}
f(x) = c - bx + \frac{1}{2} x^T A x,
\label{min_equation}
\end{equation} 
where $c$ is a constant. The gradient of the Eq. (\ref{min_equation}) is 0 when the same equation is at its minima
\begin{equation}
\nabla f(x) = Ax -b = 0.
\end{equation}
That is, at the position of the minimum of the function $f(x)$ we find a solution to Eq. (\ref{linear_system}). In most cases, finding the minimum of Eq. (\ref{min_equation}) is much easier than finding the solution of the system in \ref{linear_system}, especially when no exact solution exists for \ref{linear_system} or $A$ does not have an inverse.

The biconjugate gradient finds the minimum of Eq. (\ref{min_equation}) (or equivalently, the solution of Eq. \ref{linear_system}) by following an iterative process that minimizes the function $f(x)$ in a series of steps no longer than the dimension of the problem. The beauty of the algorithm is that the successive minimizations are carried out on a series of orthogonal conjugate directions, $p_{\rm k}$, with respect to the metric $A$. That is,
\begin{equation}
p_{\rm i } A p_{\rm j} = 0 \quad \quad j<i.
\end{equation}
This condition is useful when minimizing in a multidimensional space because it guarantees that successive minimizations do not spoil the minimizations in previous steps.

By comparising Eq. (\ref{res_square}) and Eq. (\ref{min_equation}), it is easy to identify the terms, $c = (1/2)\theta^{\rm T}\theta$, $b = \Gamma^{\rm T}$ and $A = \Gamma^{\rm T}\Gamma$. Minimizing the quantity $R^2$ is equivalent to solving Eq. (\ref{residual}). To see this, we only have to realize that
\begin{equation}
b - AX = \mathbf{\Gamma}^T (\mathbf{\Phi}-\mathbf{\Gamma X}) = \mathbf{\Gamma}^T \mathbf{R}.
\end{equation}
If an exact solution for Eq. (\ref{residual}) does not exist, the minimum of $R^2$ will be a more accuratly approximated solution to the system. The minimum can be found easily: in the case of symmetric matrices $A$, the algorithm constructs two sequences of vectors $r_{\rm k}$ and $p_{\rm k}$ and two constants, $\alpha_{\rm k}$ and $\beta_{\rm k}$. To begin the algorithm, we need to make a first guess of the solution, namely $X_0$ and two vectors $r_0$ and $p_0$

\begin{eqnarray}
\alpha_{\rm k} = \frac{r_{\rm k}^Tr_{\rm k}}{p_{\rm k}^TAp_{\rm k}}, \\
r_{\rm k+1} = r_{\rm k} - \alpha_{\rm k} A p_{\rm k}, \label{res_step}\\
\beta_{\rm k} = \frac{r_{\rm k+1}^Tr_{\rm k+1}}{r_{\rm k}^T r_{\rm k}}, \\
p_{\rm k+1} = r_{\rm k+1} + \beta p_{\rm k}
\end{eqnarray}
At every iteration, an improved estimate of the solution is given by
\begin{equation}
X_{\rm k+1} = X_{\rm k} + \alpha_{\rm k}\beta_{\rm k}.
\label{improved_estimation}
\end{equation}   
The algorithm starts with an initial guess for the solution, $X_1$, and chooses the residual and the new search direction in the first iteration to be
\begin{equation}
r_1 = p_1 = b - AX_1.
\end{equation}
We note that $p_1$ is nothing but $\nabla R^2$. Thus, the algorithm chooses as a first minimization direction the gradient of the function to be minimized at the position of the first guess. It then minimizes in directions that are conjugate to the previous ones until either it reaches a minimum or the square of the residual $R^2$ is smaller than $\epsilon$. 
 
\subsection{Quadratic programming algorithm (QADP)}\label{Appendix_QADP} 
The nonnegative quadratic programming algorithm used in this work has the peculiarity that it finds solutions, $X$, satisfying the condition $X \geq 0$. That is, negative masses are not allowed in the solution by construction. We follow the multiplicative updates proposed by \citet{Sha_Saul_Lee}. \\
The basic problem we wish to solve is to minimize the quadratic function
\begin{equation}
F({\mathbf v}) = \frac{1}{2}{\mathbf v}^{\rm t} \mathbf{Av+ b^t v},
\label{quadratic_function}
\end{equation}
subject to the constraint $v_i \geq 0, \forall i$. In Eq. (\ref{quadratic_function}), the vector ${\mathbf v}$ is the unknown vector ${\mathbf X}$, $\mathbf{A} = {\mathbf \Gamma}^{\rm T} {\mathbf \Gamma}$ and ${\mathbf b} ={\mathbf \Gamma}^{\rm T} {\mathbf \Phi} $. We note that the elements of ${\mathbf X}$ are all positive, since the $\beta$s can be chosen all positive with respect to an appropriate system of reference. The matrix {\bf A} can be decomposed into its positive and negative parts: $\mathbf{A} = \mathbf{A}^+ - \mathbf{A}^-$, where $A_{\rm ij}^+ = A_{\rm ij}$ if $A_{\rm ij} >0 $ and 0 otherwise and $A_{\rm ij}^- = -A_{\rm ij}$ if $A_{\rm ij} <0 $ and 0 otherwise ({\it nonnegative} matrices). The solution is iteratively updated by the rule
\begin{equation}
v_{\rm k+1,i} = v_{\rm k,i}\delta_{\rm i},
\label{iterative_updates}
\end{equation}
where the updating term is defined as
\begin{equation}
\delta_{\rm i} = \frac{-b_{\rm i} + \sqrt{b_{\rm i}^2+4(\mathbf{A}^+\mathbf{v})_{\rm i}(\mathbf{A}^-\mathbf{v})_{\rm i}}}{2(\mathbf{A}^+\mathbf{v})_{\rm i}}.
\label{update}
\end{equation}

It is easy to see that generic quadratic programming problems have a single unique minimum. We denote as $\mathbf v^*$ this global minimum of $F(v)$. We attempt to prove that convergence of the iteration Eq. (\ref{update}) corresponds to this minimum $\mathbf v^*$. At this point, one of two conditions must apply for each component $v^*_{\rm i}$: either (i) $v^*_{\rm i} >0$ and $\partial F / \partial v_{\rm i} (v^*_{\rm i}) = 0$ or (ii) $v^*_{\rm i}=0$ and $\partial F / \partial v_{\rm i} (v^*_{\rm i})\geq 0$. Now since
\begin{equation}
\frac{\partial F}{\mathbf v^*}=(\mathbf{A}^+\mathbf{v})_{\rm i}- (\mathbf{A}^-\mathbf{v})_{\rm i} + b_{\rm i},
\end{equation}
the multiplicative updates in both cases (i) and (ii) take the value $\delta_{\rm i} = 1$, where the minimum is a fixed point. Conversely, a fixed point of the iteration must be the minimum $\mathbf v^*$.\\


\end{document}